\documentclass{aa}  

\usepackage{graphicx}
\usepackage{txfonts}
\usepackage{lipsum}
\usepackage{subcaption}         
\usepackage{ulem}                                
\usepackage{lscape}             
\usepackage{placeins}           
\usepackage[colorlinks,linkcolor=blue,citecolor=blue]{hyperref}

\makeatletter
\AtBeginDocument{%
  \def\makeLineNumber{}%
}
\makeatother

\begin{document}

   \title{Identification and characterization of 15265 super-Nyquist frequencies in 1309 $\delta$~Scuti stars from \textit{Kepler} photometry} 

   %

   %
\titlerunning{Super-Nyquist frequencies in $\delta$ Sct stars}
   \authorrunning{Mo et al.}

   %
   \author{Yanqi Mo\inst{1,2}
      \and
      Weikai Zong\inst{1,2}\fnmsep\thanks{Corresponding author: weikai.zong@bnu.edu.cn} 
      \and
      Xuan Wang\inst{3}
      \and
            Simon Murphy\inst{4}
      \and
            Zilu Yang\inst{1,2}
      \and
            Jian-Ning Fu\inst{1,2}
      \and
            Stephane Charpinet\inst{3}
      \and
            Xiao-Yu Ma\inst{5}
   }

\institute{
    Institute for Frontiers in Astronomy and Astrophysics, Beijing Normal University, Beijing 102206, China
    \and
    School of Physics and Astronomy, Beijing Normal University, Beijing 100875, PR China
    \and
    IRAP, CNRS/Université de Toulouse/CNES, 14 Avenue Edouard Belin, Toulouse, 31400, France
    \and
     Centre for Astrophysics, University of Southern Queensland, Toowoomba, QLD 4350, Australia
     \and
     SpaceSciences, Technologies and Astrophysics Research (STAR) Institute, Université de Liège, Allée du 6 Août 19C, 4000 Liège, Belgium
   }

   \date{Received date / Accepted date}

   \abstract{
   The frequency of pressure (p) mode in $\delta$~Scuti stars can exceed the Nyquist limit of \textit{Kepler} long-cadence photometry. {These 'super-Nyquist frequencies' (SNFs) are observed as 'reflected' peaks at lower frequencies, i.e., they are Nyquist aliases that pose} a threat to asteroseismic diagnostics. Their impact on $\delta$~Scuti p modes has yet to be comprehensively explored. We performed a systematic survey to search for SNFs in 1,838 \textit{Kepler} $\delta$~Scuti stars through a novel technique based on sliding Lomb-Scargle periodogram, identifying 15,265 confirmed SNFs in 1,309 stars, from a total of 259,883 frequencies.
   We observe that the total number of detected frequencies per star remains featureless across the $\delta$~Scuti instability strip; however, young stars pulsate in higher frequencies and so have significantly more SNFs on average. Both the number and the rate of SNFs diminishes accordingly as $\delta$~Scuti stars become more evolved, which is consistent with both observation and stellar models.
   Furthermore, our method detects a greater fraction of modes as SNFs at higher frequencies, rising from approximately 1\% at 20 \(\mu \)Hz to 23\% at the Nyquist limit. The rate of underdetection is highest amongst low-amplitude modes. The SNF modulation patterns can be well distinguished from phase modulations induced by binarity or nonlinear mode interactions. We provide a frequency catalog for future asteroseismic studies of $\delta$~Scuti stars, wherein we identify each peak as being real or an alias, enabling further investigations into regular patterns of pulsation modes, linear combination frequencies, and theoretical modeling.
}


   \maketitle

\section{Introduction}
\label{sec:intro}

Asteroseismology, a unique technique for probing the internal structures of stars through pulsation modes, is now applied to various types of stars across the entire Hertzsprung-Russell diagram \citep[see, e.g.,][]{2021RvMP...93a5001A,2022ARA&A..60...31K}. As one of the popular classes of pulsators, $\delta$~Scuti stars are numerous and typically exhibit relatively high amplitudes.
These intermediate-mass stars typically have masses \(1.5\)--\(2.5\,\mathrm{M}_\odot\) and effective temperatures $T_\mathrm{eff}\simeq6400$--$8600$~K, lying at the intersection of the classical instability strip and the main sequence \citep{2000ASPC..210....3B,2001A&A...366..178R}.
Their pulsations are driven by the $\kappa$-mechanism operating in the He~II ionization zone,
which primarily excites low-order pressure (p) modes with typical periods ranging from $\sim$15~min to 5~h.
In addition to these p modes, many $\delta$~Sct stars also exhibit low-frequency gravity (g) modes and, in more evolved objects, mixed modes that display g-mode sensitivity near the core and p-mode characteristics near the surface \citep{1975PASJ...27..237O,2011MNRAS.417..591B}.

Before space photometry, the discovery of $\delta$~Scuti stars from ground-based observations led to progress in mode identification and stellar modeling, although it was hindered by daily aliasing and poor frequency resolution \citep[see, e.g.,][]{1988Ap&SS.140..255W,1998A&A...332..958B,1999MNRAS.302..349B,2001A&A...366..178R}. These photometric limitations were partially ameliorated by observations from multi-site networks or Antarctica \citep[see, e.g.,][]{2000MNRAS.318..511H,2005A&A...435..955B,2015AJ....149...84Z}, which provided richer pulsation spectra with improved frequency resolution. Subsequently, spaceborne photometry initiated a new era for the asteroseismology of $\delta$~Scuti stars, revealing a wealth of low-amplitude frequencies with sharp frequency resolution \citep[see, e.g.,][]{2009A&A...506...85P,2014MNRAS.437.1476B}.

The {Kepler} and Transiting Exoplanet Survey Satellite (TESS) missions have collected high-quality photometry for thousands of $\delta$~Scuti stars \citep{2010Sci...327..977B,2011A&A...534A.125U,2015JATIS...1a4003R,2024ApJ...972..137G,2025Univ...11..302Z},
brought advances in linear asteroseismology regarding pulsation properties. For example, $\delta$~Scuti stars have been found to exhibit regular frequency spacings \citep{2011Natur.477..570A,2020Natur.581..147B}, possess a remarkably high fraction of additional low-frequency g modes \citep{2014MNRAS.437.1476B}, and reveal statistical relations in period-luminosity \citep{2022MNRAS.516.2080B} and frequency patterns dependent on stellar parameters \citep{2018MNRAS.476.3169B,2018A&A...614A..46B}. Their instability strip can now be better constrained through observational efforts \citep{2019MNRAS.485.2380M,2024ApJ...972..137G}.
Furthermore, their frequency properties are valuable for constraining physical processes involving stellar rotation \citep{2021MNRAS.505.6217R}, magnetic fields \citep{2015MNRAS.454L..86N,2020A&A...643A.110Z}, and tidal interactions \citep{2020NatAs...4..684H,2021MNRAS.503..254R}. These stars may reside in binary systems \citep{2022ApJS..263...34C}, which is essential for understanding mass-transfer processes \citep{2017ApJ...837..114G}, or be located in open clusters allowing for precise age dating \citep{2022MNRAS.513..374P,2023ApJ...946L..10B}. Moreover, artificial intelligence has been introduced into $\delta$~Scuti studies for tasks such as light curve identification \citep{2022MNRAS.514.2793B}, emulating model grids \citep{2023MNRAS.525.5235S}, and classifying island modes \citep{2019MNRAS.483L..28M}.

Beyond these advances in pulsation properties, \textit{Kepler} and TESS opened new avenues for characterizing temporal variations of pulsation amplitude and phase. \citet{2012MNRAS.422..738S} proposed a theoretical framework for detecting orbital companions via frequency modulation, and determining their orbital parameters \citep{2015MNRAS.450.3999S}, which was subsequently adapted to discover hundreds of new binaries via pulsation timing in \textit{Kepler} A/F pulsators \citep{2018MNRAS.474.4322M}. Regarding amplitude variation, \citet{2014ApJ...783...89B} and \citet{2015A&A...579A.133B} observed clear correlations between three interacting modes, interpreting them as nonlinear resonant couplings. This phenomenon was found to be common following a survey of large-amplitude pulsations in 983 $\delta$~Scuti stars \citep{2016MNRAS.460.1970B}. Notably, geometric effects in binary systems can induce amplitude modulation of $\delta$~Scuti pulsations, over short timescales \citep[see, e.g.,][]{2020MNRAS.494.5118K}. Recently, amplitude and phase modulations have gained increasing attention in $\delta$~Scuti studies, as the diversity in modulation patterns remains not fully understood \citep[see, e.g.,][]{2023AJ....166...43N}.

Crucially, some pulsation modes in $\delta$~Scuti stars can exceed the Nyquist frequency of the \textit{Kepler} long-cadence photometry \citep{2013MNRAS.430.2986M}.
In this regime, reflected alias signals show characteristic amplitude and phase modulation caused by the sampling process \citep{2014MNRAS.441.2515M,2016MNRAS.460.1970B}.
These high frequencies, when aliased to lower frequencies by the Nyquist limit, are termed super-Nyquist frequencies (SNFs).
The modulated time sampling of \textit{Kepler} data introduces distinctive multiplet patterns in the periodograms of reflected frequencies \citep{2013MNRAS.430.2986M}, while barycentric time corrections induce periodic amplitude and frequency modulations that serve as diagnostic features of SNFs \citep{2021RNAAS...5...41Z}.
Leveraging these characteristics, \citet{2025A&A...693A..63W} proposed a time-dependent analysis to identify SNFs using the sliding Lomb-Scargle periodogram (sLSP).
Their simulations confirmed the feasibility and computational efficiency of this method, subsequently applying it to 611 $\gamma$~Doradus stars, which raised a cautionary note that hybrid p modes could be aliased frequencies mixed with SNFs up to $\sim7$\%. Following this, \citet{2025Univ...11..246Y} extended the approach to $\delta$~Sct stars, analyzing 1400 frequencies in 68 targets from the literature and distinguishing six previously unknown reflected SNFs.

However, a systematic survey of SNFs across the entire $\delta$~Sct population has yet to be explored. To address this, we apply the sLSP technique to 1,838 $\delta$~Sct stars from the prime \textit{Kepler} photometry, based on an automated implementation to upgrade the sLSP workflow.
The structure of this paper is organized as follows: Section~\ref{sec:method} describes the data and the automated pipeline used to identify SNFs. Section~\ref{sec:results} presents the main results of the survey, including the statistical distribution and physical implications of SNFs in $\delta$~Sct stars. Finally, Section~\ref{sec:discussion} discusses these findings and summarizes the main conclusions of this work.

\section{Data and Methods}
\label{sec:method}

\begin{figure*}[htb!]
    \centering
    \includegraphics[width=\textwidth]{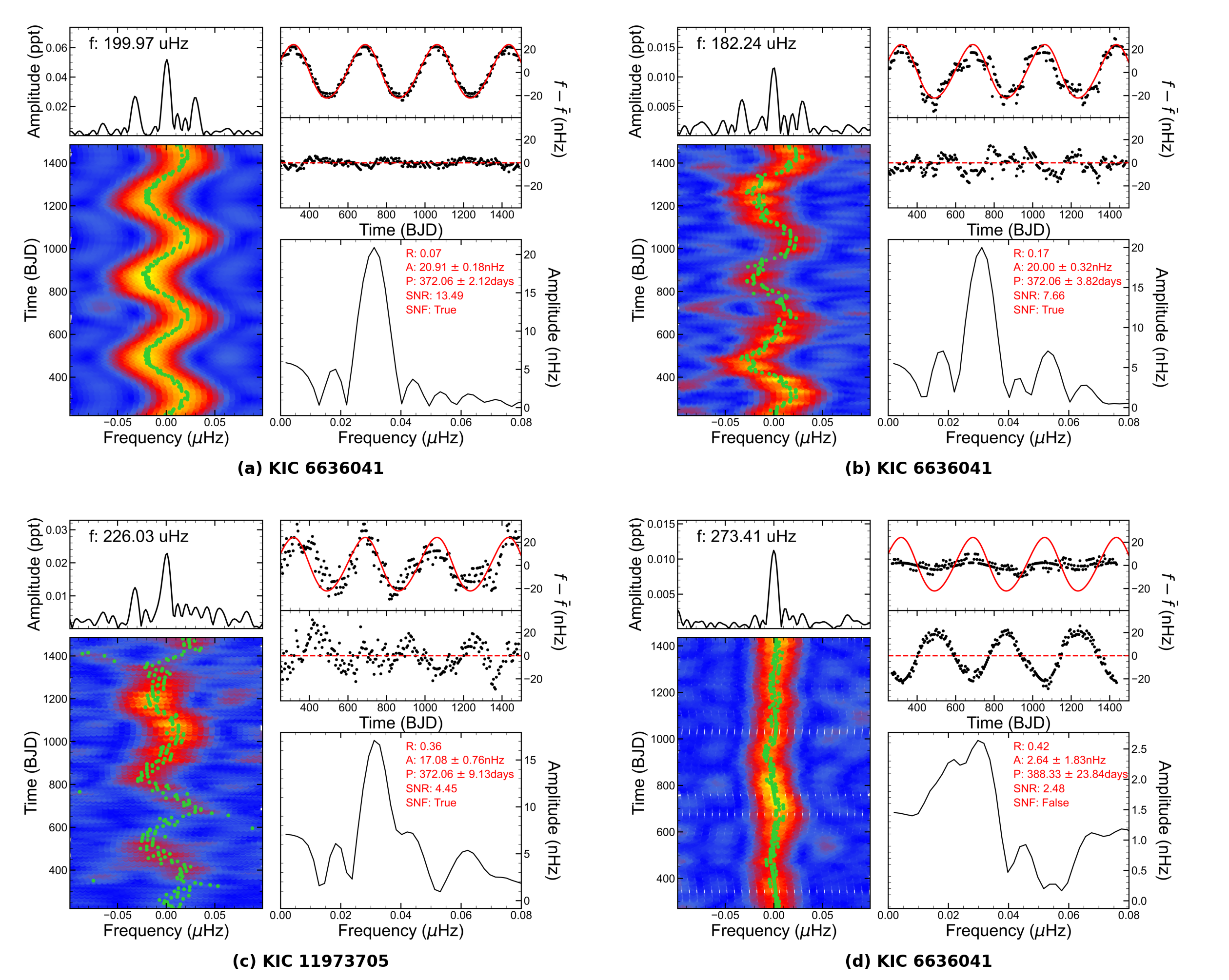}
    \caption{Representative diagnostic plots generated by our automated pipeline. 
    The classification results of four distinct types of candidate frequencies are shown as follows: panels (a), (b), and (d) correspond to KIC\,6636041, while panel (c) corresponds to KIC\,11973705. Panels (a)-(c) are classified as SNFs, whereas panel (d) is classified as not an SNF.
    Each diagnostic plot displays a composite of five subplots:
    (\textit{top left}) The local LSP with the target frequency marked.
    (\textit{bottom left}) A sliding Lomb-Scargle periodogram (sLSP) showing the time-dependent frequency behavior, with amplitude indicated by the color. The green dots mark the frequency of maximum amplitude within each LSP segment centered around that frequency at given time.
    (\textit{top right}) The modulated frequency time series $f(t)$ (black points) compares with the theoretical SNF modulation pattern derived from the spacecraft orbit (red curve). The residuals of the fit are shown in the {\it middle right}, marked with a dashed horizontal line at zero.
    (\textit{bottom right}) The LSP of $f(t)$, annotated with the key diagnostic parameters used for classification: the residual variance ($R$), modulation amplitude ($A_{\rm mod}$), modulation period ($P_{\rm mod}$), the modulation signal-to-noise ratio (SNR), and the final automated classification flag (True or False).    
    }
    \label{fig:slsp_examples}
\end{figure*}

The {Kepler} Mission provides high-precision, nearly continuous photometry that is ideal for $\delta$~Sct studies.
Observations are organized into quarters (Q0–Q17), which were typically about three months long and separated by $\sim$90$^\circ$ spacecraft rolls, with some differences at the beginning and end of the mission.
The primary mission provided two observing cadences: long cadence (LC; $\sim$29.4\,min) and short cadence (SC; $\sim$58.9\,s) \citep{2010Sci...327..977B}.

In this paper, we use Q to denote these observing segments and require a sufficiently long time baseline for robust frequency analysis.
We constructed our sample based on two foundational catalogs: the 983 $\delta$~Sct stars in \citet{2016MNRAS.460.1970B} and 1,988 stars in \citet{2019MNRAS.485.2380M}. 
After cross-matching and removing duplicates, we retained only those stars with light curves spanning more than 14 \textit{Kepler} quarters to ensure sufficient frequency resolution and time baseline, yielding a final sample of 1,838 stars.

Photometric data for all targets were retrieved from the Mikulski Archive for Space Telescopes (MAST)\footnote{\url{http://archive.stsci.edu/kepler/}}. 
We primarily utilized the Pre-search Data Conditioning Simple Aperture Photometry (PDCSAP; \citealt{2012PASP..124.1000S, 2012PASP..124..985S}) light curves, 
which were processed by the NASA {Kepler} Data Processing Pipeline \citep{2010ApJ...713L..87J}. 
These data were obtained in long-cadence (LC) mode with a typical sampling interval of $\Delta t \simeq 29.4$ minutes, 
corresponding to a Nyquist frequency of $f_{\rm ny} \approx 283\,\mu\mathrm{Hz}$.
Data processing was performed following the procedures outlined in \citet{2024ApJS..271...57X} and \citet{2025A&A...693A..63W}. 
The light curves were initially cleaned and normalized using the \texttt{lightkurve} package \citep{2018ascl.soft12013L}. 
For a subset of stars exhibiting poor data quality or significant contamination in the standard apertures, 
we downloaded the corresponding Target Pixel Files (TPFs). 
For these cases, we defined custom apertures using a threshold mask within the \texttt{lightkurve} package \citep{2018ascl.soft12013L} and re-extracted the light curves, see detailed processes in \citet{2023A&A...680A..11M}.
To remove instrumental systematics and stitch the observational quarters, the light curves were detrended using the \texttt{flatten} routine from the \texttt{wotan} package \citep{2019AJ....158..143H}. 
This step effectively removed residual long-term trends while preserving the intrinsic stellar variability. 
Finally, the cleaned and detrended light curves were analyzed using the \texttt{FELIX} software \citep{2010A&A...516L...6C, 2016A&A...585A..22Z} to extract a comprehensive list of significant pulsation frequencies.

The barycentric time correction applied to \textit{Kepler} photometry causes the Nyquist frequency to vary periodically with the orbital motion of \textit{Kepler}. 
This effect introduces a predictable modulation pattern in the frequencies above $f_{\rm ny}$, a signature addressed by \citet{2013MNRAS.430.2986M}.
To capture this behavior, the sliding Lomb--Scargle periodogram (sLSP) divides the light curve into successive, overlapping time windows, computes a periodogram in each window, and then tracks the temporal evolution of each candidate peak in frequency and amplitude. Reflected SNFs are identified by their nearly periodic modulation patterns in these time series. 
The challenge is particularly acute in $\delta$~Scuti stars given their dense frequency spectra. 
Accordingly, we develop \texttt{sLSP4SNFs}\footnote{\url{https://github.com/XuanWang-star/sLSP4SNFs}}, a robust and systematic pipeline, dedicated to the automated identification of SNFs via signatures characterized by \citet{2025A&A...693A..63W}.

The pipeline processes each star through a standardized workflow. 
First, candidate frequencies are selected from the \texttt{FELIX} list with a threshold ${\rm SNR} > 8$. 
The nearby peaks are then consolidated using a grouping algorithm that retains only the highest-SNR component within the frequency resolution. 
For stars with exceptionally dense spectra, the analysis is restricted to the 1,024 strongest frequencies. 
Each candidate is analyzed using a sliding LSP with a 200-day window (moving in 5-day steps) to produce a time-dependent frequency $f(t)$.
This window/step configuration is chosen as a practical compromise between frequency resolution and computational cost: the 200-day window provides stable local frequency estimates, while the 5-day step samples modulation with sufficient temporal density.
This setup is guided by the expected SNF modulation induced by barycentric time correction \citep{2013MNRAS.430.2986M,2025A&A...693A..63W}. Over the four-year \textit{Kepler} baseline, reflected SNFs are expected to show repeating modulation cycles, providing a practical diagnostic for identification.
From this, we compute key diagnostic parameters: the modulating period ($P_{\rm mod}$), amplitude ($A_{\rm mod}$), signal-to-noise ratio (${\rm SNR}_{\rm mod}$), and the residual, $R$, which quantifies the goodness-of-fit between the observed $f(t)$ and the theoretical SNF modulation.

We adopted a hierarchical classification strategy to identify valid SNFs after testing the robustness of the \texttt{sLSP4SNFs} pipeline. We applied it to all 45,607 frequencies in $\gamma$~Dor stars extracted by \citet{2025A&A...693A..63W}, which returns 782 SNF candidates. After cross-matching with the 304 SNFs from their visual inspection, candidates who exhibit an exceptionally strong match to the theoretical expectation, defined by $R < 0.3$, were automatically accepted. 
For candidates with slightly higher residuals ($0.3 \le R < 0.5$), we imposed stricter criteria requiring significant modulation significance (${\rm SNR}_{\rm mod} > 3$) and a modulation period consistent with the \textit{Kepler} orbital year ($P_{\rm mod} \approx 372$\,d). 
If a candidate failed to meet these criteria using the 200-day window, the entire analysis was repeated using a 300-day window to enhance frequency resolution for weaker signals.

The \texttt{sLSP4SNFs} automatically generates a comprehensive diagnostic plot for each candidate, summarizing the classification metrics, and outputs a catalog of quantitative parameters.
We applied this pipeline to our full sample of 1,838 $\delta$~Scuti stars. 
Figure~\ref{fig:slsp_examples} displays representative diagnostic plots ordered by signal quality, ranging from high-confidence detections to clear non-detections. 
For frequencies with high SNR$_\mathrm{mod}$, we clearly observe that our measured frequency modulation follows a pattern consistent with the numeric expectation, resulting in minimal residuals.
However, in some cases, a frequency may exhibit intrinsic modulation superimposed on the SNF pattern, which requires further inspection, for instance, the frequency around 251.8\,$\mu$Hz in KIC~3852644 (Fig.~\ref{fig:slsp_examples}c).
A systematic review of these diagnostic plots indicated that candidates in the marginal regime of ${\rm SNR}_{\rm mod} \in (3,~5)$ frequently exhibited ambiguous signatures. 
Consequently, we implemented a final manual vetting step for all candidates in this specific SNR range to ensure the reliability of the final catalog, whereas candidates with lower SNR$_\mathrm{mod}$ were automatically discarded.

\section{Results}
\label{sec:results}

\begin{table*}
\caption{The SNF Catalog for 1,306 $\delta$~Scuti Stars.}
\label{tab:snf_catalog_excerpt}
\centering
\begin{tabular}{c c c c c c c c c c c} 
\hline\hline 
KIC & Frequency & $\sigma_{\rm f}$ & Amplitude & $\sigma_{\rm A}$ & $\mathrm{SNR}_{\rm FELIX}$ & $\mathrm{SNR}_{\rm mod}$ & $f_{\rm R}$ & C & Range & SC \\
 & ($\mu$Hz) & ($\mu$Hz) & (ppt) & (ppt) & & & ($\mu$Hz) & & & \\
\hline 
1026294 & 258.409096 & 0.000216 & 0.0108 & 0.0005 & 20.1 & 6.0 & 308.0149038 & 0 & 1 & -9 \\
1026294 & 250.283514 & 0.000007 & 0.4032 & 0.0006 & 653.6 & 16.4 & 316.1404861 & 0 & 1 & -9 \\
1026294 & 247.072167 & 0.000037 & 0.0696 & 0.0006 & 117.5 & 15.8 & 319.3518327 & 0 & 1 & -9 \\
1026294 & 229.689858 & 0.000014 & 0.1866 & 0.0006 & 312.5 & 12.3 & 336.7341418 & 0 & 1 & -9 \\
1026294 & 222.255286 & 0.000019 & 0.1288 & 0.0006 & 226.1 & 15.4 & 344.1687140 & 0 & 1 & -9 \\
1026294 & 218.260433 & 0.000136 & 0.0176 & 0.0006 & 32.0 & 10.5 & 348.1635669 & 0 & 1 & -9 \\
1026294 & 206.899067 & 0.000161 & 0.0158 & 0.0006 & 27.0 & 13.6 & 359.5249334 & 0 & 1 & -9 \\
\dots & \dots & \dots & \dots & \dots & \dots & \dots & \dots & \dots & \dots & \dots \\
\hline 
\end{tabular}

\tablefoot{
Parameters include the KIC identity, pulsation frequency, and amplitude with their respective uncertainties ($\sigma_{\rm f}$, $\sigma_{\rm A}$), along with the detection significance ($\mathrm{SNR}_{\rm FELIX}$) derived from \texttt{FELIX}. 
$\mathrm{SNR}_{\rm mod}$ represents the significance of the frequency modulation detected by our automated sLSP analysis.
The real frequency above the Nyquist limit is denoted as $f_{\rm R}$, and $C$ is the classification flag, indicating a combination frequency (1) or an independent frequency (0). 
In the column \texttt{Range}, the Nyquist interval is inferred from amplitude ratio method: integer $n\in[1,4]$ correspond to ranges above the Nyquist frequency of 
$[nf_{\rm ny}, (n+1)f_{\rm ny}]$, and -9 indicates no available determination from the amplitude ratio.
The column \texttt{SC}: 1 indicates confirmation by available SC photometry, and -9 denotes no available SC photometry.
See details in the text. The full catalog of 15,625 SNFs is available in machine-readable form in the online version.
}
\end{table*}

\begin{figure}[htb ]
    \centering
    \includegraphics[width=\columnwidth]{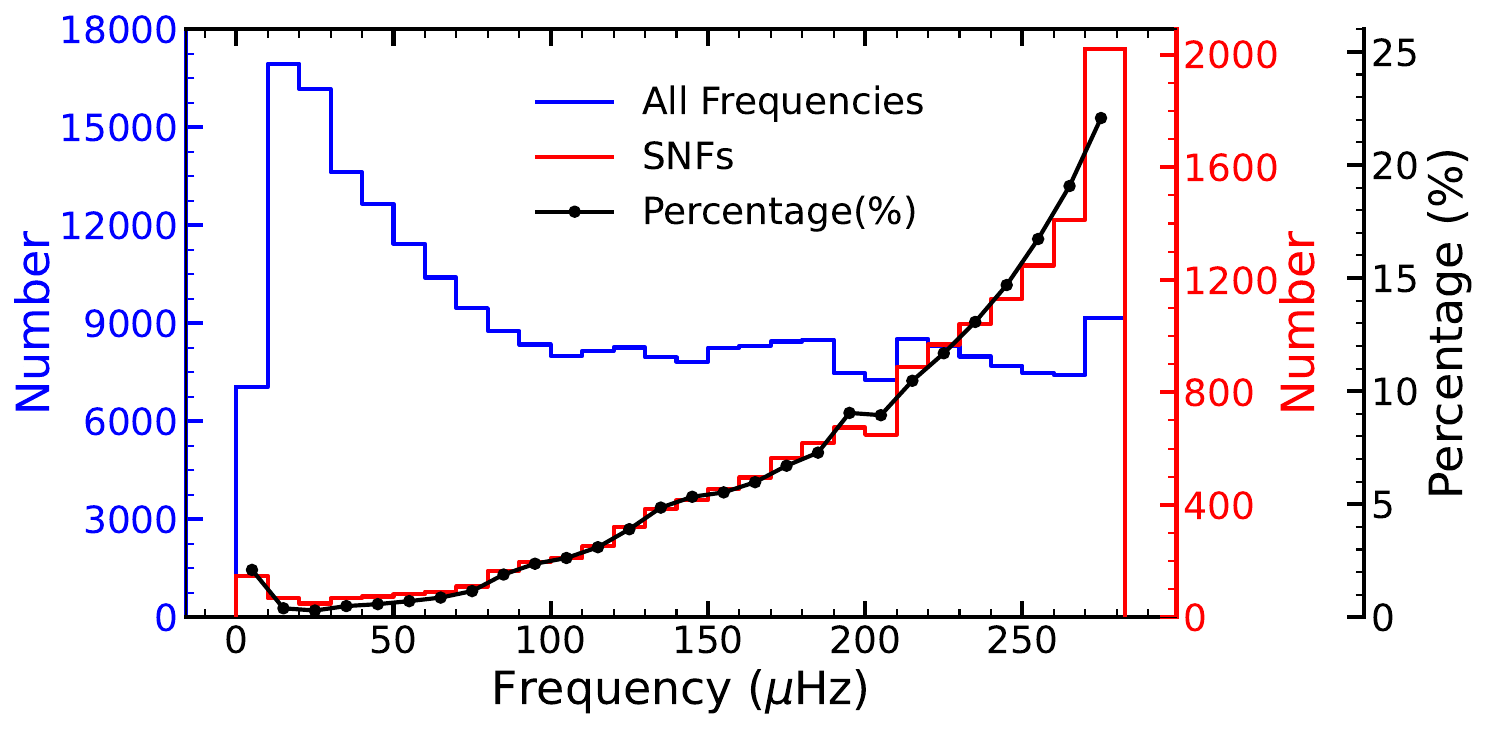}
    \caption{
    Distribution of all 259,883 frequencies in 1,838 $\delta$~Scuti stars.
    The blue histogram represents all frequencies, while the red one corresponds to the 14,824 SNFs. The black dots indicate the percentage of SNFs relative to all detected frequencies.
    }
    \label{fig:freq_distribution}
\end{figure}

Our automated pipeline initially identified 14,976 super-Nyquist frequency (SNF) candidates from a total of 259,883 candidate frequencies across 1,838 $\delta$~Scuti stars. 
As established in Section~\ref{sec:method}, candidates in the low-SNR regime require manual vetting. 
Accordingly, all 2,862 candidates with a modulation signal-to-noise ratio (${\rm SNR}_{\rm mod}$) below 5 were subjected to visual inspection. 
This process led to the rejection of 152 unreliable detections, resulting in a high validation rate of 94.7\% for our automated low-SNR classification. 
This yields a final, robust catalog of 14,824 confirmed SNFs in 1,291 stars, demonstrating that SNFs are a common feature, present in 70.2\% of the sample. 
Representative examples of the SNF catalog entries are listed in Table~\ref{tab:snf_catalog_excerpt}, while the complete catalog is available in a machine-readable format online.

Figure~\ref{fig:freq_distribution} displays the frequency distribution of all extracted frequencies and identified SNFs. We observe that the number of frequencies initially decreases as the frequency increases after peaking at $\sim$10~$\mu$Hz, which is consistent with the high density of g~modes at low frequencies in hybrid $\delta$~Scuti stars. The distribution gradually flattens around $\sim$100~$\mu$Hz; this behavior is different from that found in $\gamma$~Dor stars \citep{2025A&A...693A..63W}. 
In contrast, the number of SNFs rises monotonically with frequency, with their fraction increasing from approximately 1\% at low frequencies ($\sim$10-50~$\mu$Hz) to $\sim$23\% near the \textit{Kepler} long-cadence Nyquist limit of $f_{\rm ny}\sim$283~$\mu$Hz. 
This trend indicates that a significant portion of high-frequency p~modes reside above $f_{\rm ny}$, consistent with the finding of \citet{2019MNRAS.485.2380M}, and that recovering SNFs is essential to obtain the correct pulsation content.

\begin{figure*}[h]
    \noindent
    
    \begin{minipage}{0.65\linewidth}
        \raggedright        \includegraphics[width=0.95\textwidth]{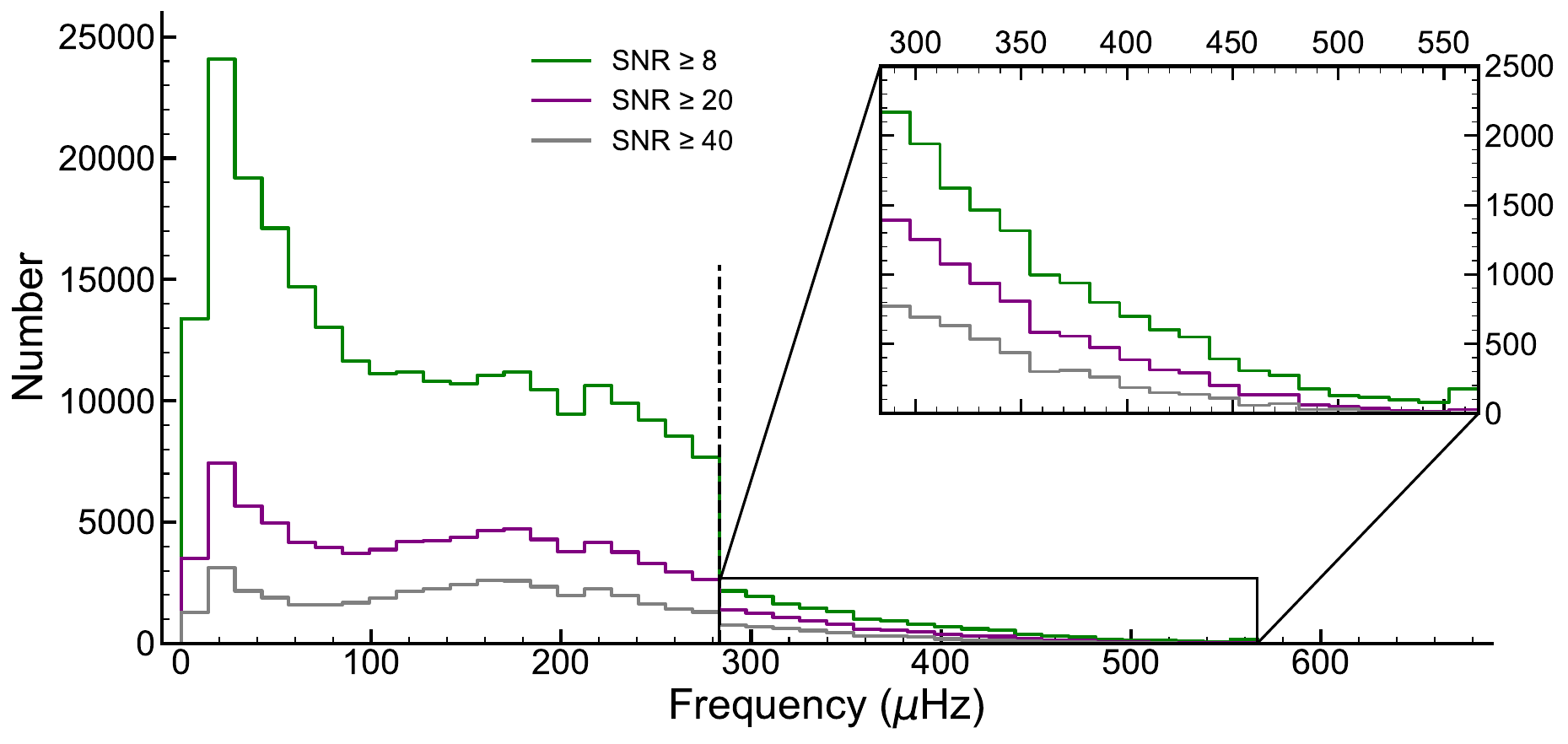}
    \end{minipage}
    \begin{minipage}{0.3\linewidth}
        \raggedleft
        \caption{Distribution of all frequencies, including recovered SNFs, in the \textit{Kepler} \(\delta\)~Scuti sample. The vertical dashed line indicates the \textit{Kepler} long-cadence Nyquist frequency, and the inset panel highlights the SNF region. Note that the discontinuity decreases as SNR increases.}
        \label{fig:snf_distribution}
    \end{minipage}
\end{figure*}

To investigate the distribution of the true pulsation frequencies, we recovered them from the observed SNFs. 
Our automated method is designed to identify the modulation signature characteristic of real frequencies, $f_{\rm R}$, restricted to the frequency range between the Nyquist frequency and twice its value ($f_{\rm ny} < f_{\rm R} < 2f_{\rm ny}$), although some pulsation frequencies can exceed 2$f_{\rm ny}$ \citep[see, e.g.,][]{2020Natur.581..147B}. This is because the robustness of SNFs in that higher frequency range has yet to be fully validated, constrained by the current lack of a sufficient known sample. Thus the real frequency is recovered from its observed alias, $f_{\rm obs}$, using the standard reflection relation:
\begin{equation}
    f_{\rm R} = 2f_{\rm ny} - f_{\rm obs}.
    \label{eq:snf_reflection}
\end{equation}
The resulting distribution of these recovered frequencies reveals a prominent discontinuity at $f_{\rm ny}$ (the mean Nyquist frequency of 283.212~$\mu$Hz), as shown in Fig.~\ref{fig:snf_distribution}.
This step-feature marks the boundary above which all identified frequencies are those recovered by our automated pipeline from their aliased signals. 
The sharp drop in density at $f_{\rm ny}$ is a consequence of the inherent incompleteness of our pipeline: if all SNFs could be recovered, a smoother distribution would be expected. 
Nevertheless, the large population of frequencies now identified above $f_\mathrm{ny}$ demonstrates the efficiency of our approach in detecting high-frequency pulsations. 
A close inspection of Fig.~\ref{fig:snf_distribution} reveals a notable trend: as the SNR threshold is raised, the number of recovered SNFs declines more slowly than that of the real frequencies below $f_{\rm ny}$. 
This suggests that the recovery rate of SNFs identified by our pipeline increases with SNR. The frequency ratio across $f_\mathrm{ny}$ exceeds 50\% at SNR=20. Our results indicate that a significant fraction of SNFs is missing among lower-SNR frequencies via our \texttt{sLSP4SNFs} code.

Based on our frequency samples, we searched for combination frequencies in both the original and recovered sets, as the recovery of SNFs may affect linear combination relationships. 
The identification process consisted of screening for harmonics ($f_n \approx n \cdot f_i$, integer $n \le 99$) to isolate independent frequencies by order of their amplitudes, followed by a search for linear combinations of two frequencies ($f_{\rm comb} = a f_i + b f_j$), where $a,b \in \{ -1,\,+1 \}$ among them. 
We accept a candidate frequency $f_k$ as a combination of independent frequencies $f_i$ and $f_j$ only if:
\begin{equation}
    |f_k - (af_i + bf_j)| \le 3\sqrt{\sigma_k^2 + \sigma_i^2 + \sigma_j^2},
\end{equation}
where $\sigma$ denotes the frequency uncertainty.

Our analysis identified 1,438 linear combination frequencies in the original extraction results and 1,733 after SNF recovery. 
The combination frequencies identified by our method are labeled with "1" in Table~\ref{tab:snf_catalog_excerpt}.
This difference of approximately 17\% between the two counts reflects the shift of aliased SNFs to their real frequencies, which is crucial for future precise seismic modeling.

\begin{figure}[htb ]
    \centering
    \includegraphics[width=\columnwidth]{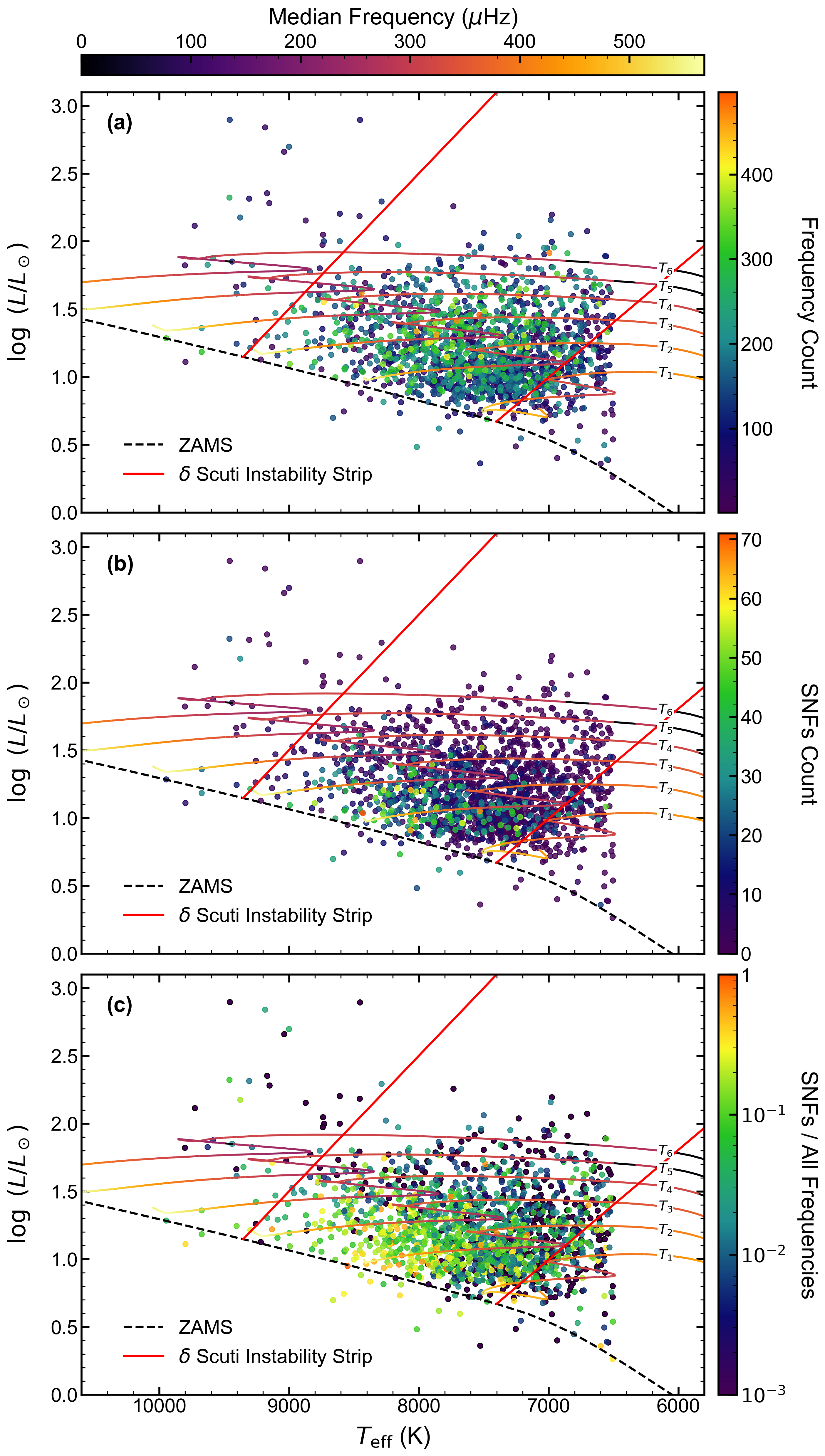}
\caption{
    HR diagrams of the 1,776 $\delta$~Scuti stars in our sample.
The data points are color-coded according to the right-hand colorbars, representing:
(a) the total number of detected frequencies;
(b) the number of recovered SNFs; and
(c) the fraction of SNFs relative to the total number of frequencies.
The solid red curves and the black dashed line mark the observational instability strip for $\delta$~Scuti stars introduced by \citet{2019MNRAS.485.2380M} and the Zero-Age Main Sequence (ZAMS), respectively.
The curves (labeled $T_1$--$T_6$ from bottom to top) show six representative evolutionary tracks with masses of $M = 1.5,~1.7,~1.9,~2.1,~2.3,~2.5\,\,\mathrm{M}_\odot$ and metallicity $Z=0.014$.
The color of these tracks indicates the median values of theoretical pulsation frequency for a given mode with fixed \(n\), \(\ell \) and \(m\), as mapped by the top colorbar.
}
    \label{fig:HR_fig}
\end{figure}

To investigate the stellar properties linked to SNF occurrence, we examined the distribution of our sample on the Hertzsprung–Russell (HR) diagram. 
We constructed stellar models using the stellar evolution code MESA \citep[Modules for Experiments in Stellar Astrophysics,][]{2019ApJS..243...10P}. A grid of 154 evolutionary tracks was computed with initial masses $M\in[1.5,~2.5]\,\mathrm{M}_\odot$ ($\Delta M=0.1\,\mathrm{M}_\odot$), metallicity $Z\in[0.004,~0.030]$ ($\Delta Z = 0.002$), and other parameters set to default values. 
Each evolutionary track consists of 300 discrete models; for each model, we calculated the theoretical frequencies for p modes with $\ell=0,~1,~2$ and $n_p=1-7$ using the stellar oscillation code GYRE \citep{2013MNRAS.435.3406T}. For comparison with observations, we selected six representative tracks with masses of $M = 1.5,~1.7,~1.9,~2.1,~2.3,~2.5\,\mathrm{M}_\odot$ and a solar metallicity of $Z=0.014$. 
Among the 1,838 stars analyzed, 1,776 have available $T_\mathrm{eff}$ and luminosity parameters provided by \citet{2019MNRAS.485.2380M}. 
Figure~\ref{fig:HR_fig}(a) shows the distribution of all stars colored by their total number of detected frequencies, revealing no clear systematic trend across the instability strip. 
In contrast, Fig.~\ref{fig:HR_fig}(b) colors the stars by their number of detected SNFs, showing a pronounced enhancement toward the hotter, less luminous region near the ZAMS. 
A similar pattern is observed in Fig.~\ref{fig:HR_fig}(c), which shows the fraction of SNFs relative to the total number of detected frequencies per star. A clear decrease in the SNF fraction is observed as stars evolve away from the ZAMS, correlating well with the median eigenfrequencies from GYRE models. These findings indicate that the prevalence of SNFs is higher in younger $\delta$~Scuti stars, rather than a higher number of detected frequencies.

\section{Discussion and summary}
\label{sec:discussion}
\begin{figure}[htb]
    \centering
    \includegraphics[width=\columnwidth]{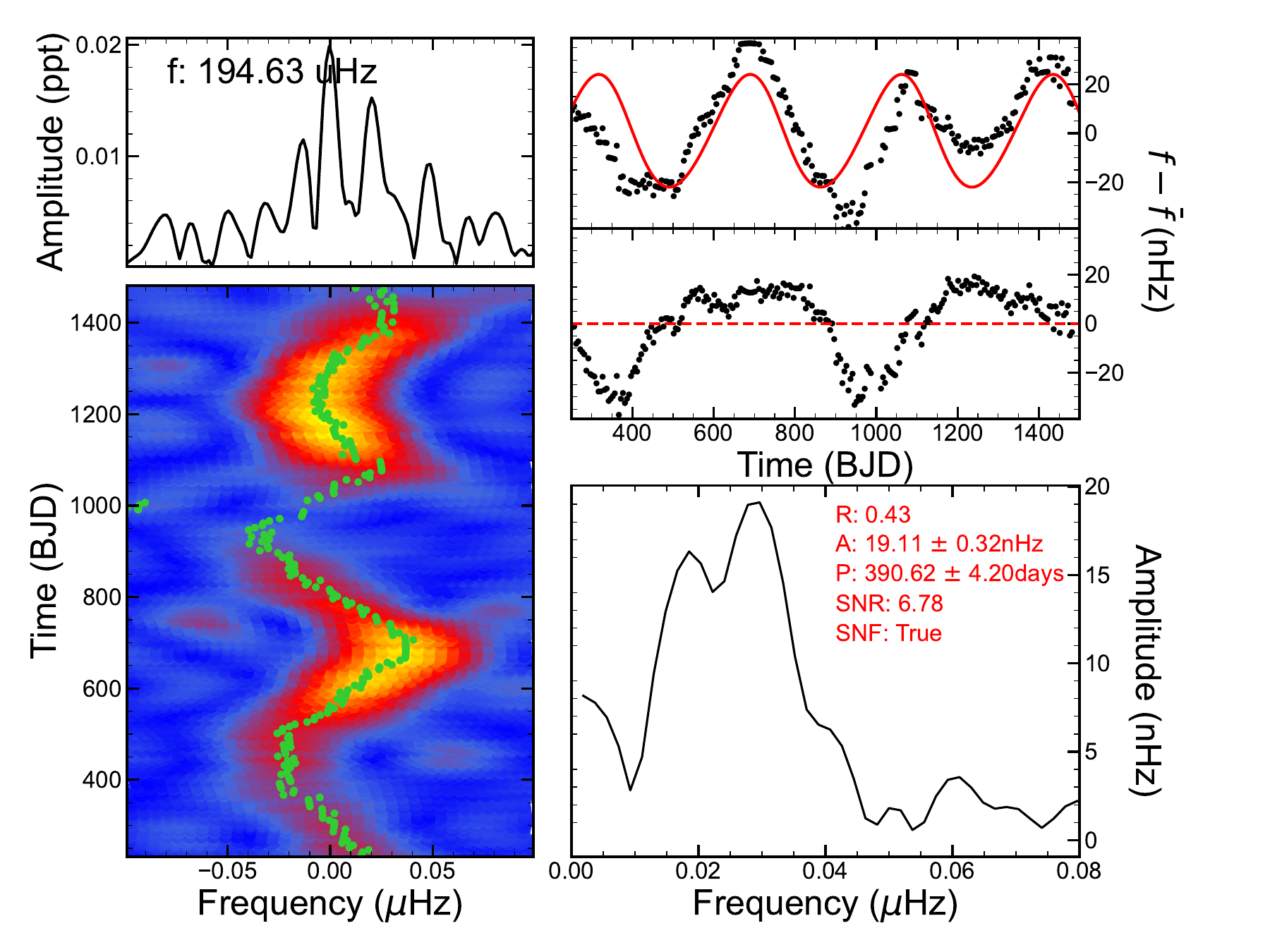}
\caption{Same as Fig.\,\ref{fig:slsp_examples} but for a potential false-alarm SNF in KIC\,7848288.}
    \label{fig:sLSPmissingSNF}
\end{figure}

\begin{figure}[htb]
    \centering
    \includegraphics[width=\columnwidth]{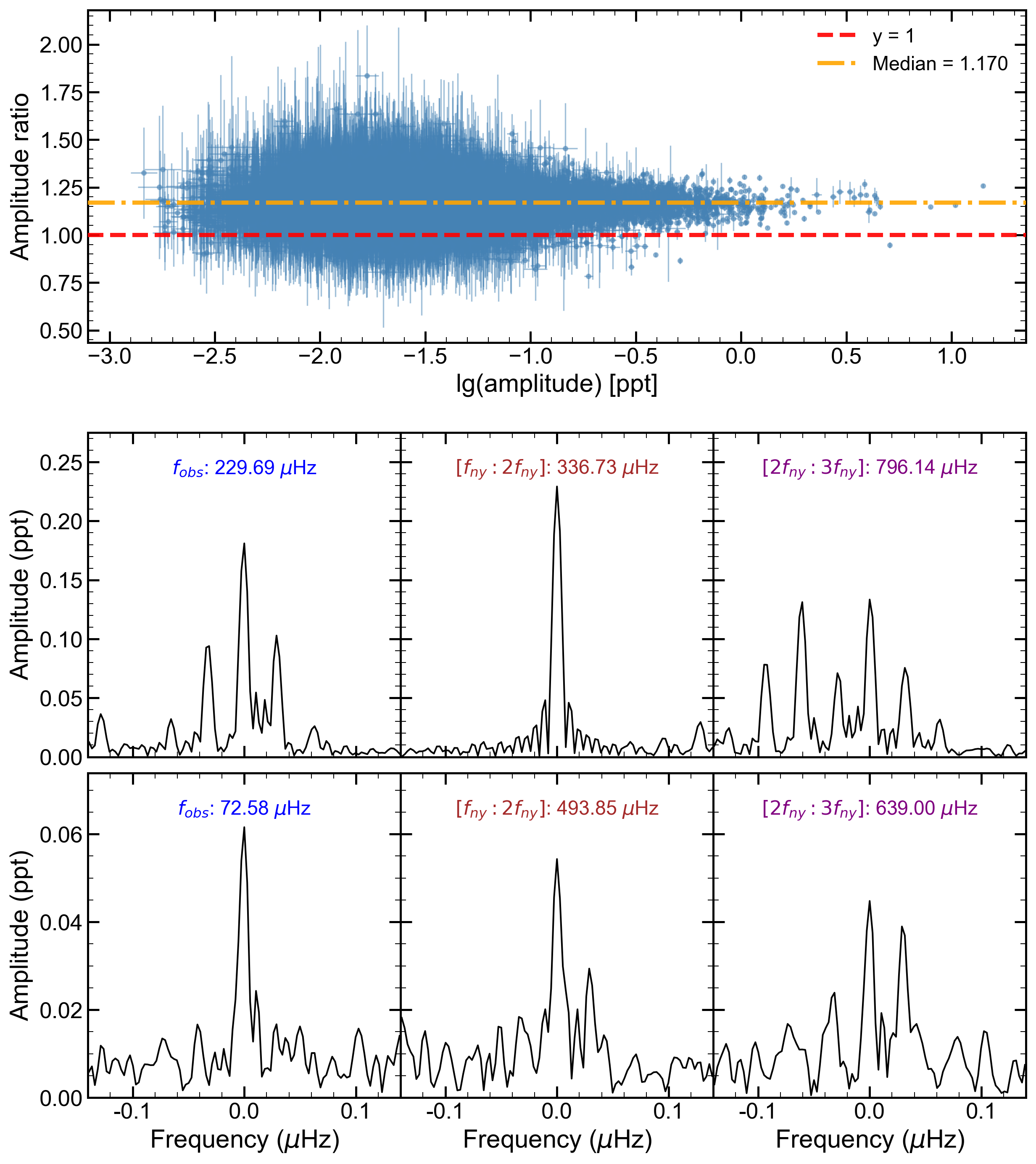}
\caption{ Validation of 14,824 SNFs via alias amplitude comparison. Top panel: Distribution of amplitude ratios for frequency aliases within the \([0,f_{\text{ny}}]\) and \([f_{\text{ny}},2f_{\text{ny}}]\) intervals, shown as a function of amplitude on a logarithmic scale. The red dashed and orange dash-dotted horizontal lines mark \(y=1\) and the median amplitude ratio, respectively. Middle panels: A representative example of one of the 14,509 confirmed candidates where the physical frequency correctly resides in the \([f_{\text{ny}},2f_{\text{ny}}]\) range. Bottom panels: An illustrative example of one of the 283 unconfirmed cases that do not satisfy this amplitude criterion. Note the text labels indicate the Nyquist-interval assignment and the corresponding recovered frequency values (e.g., \(f_{\mathrm{obs}}\) in \([f_{\text{ny}},2f_{\text{ny}}]\) and \([2f_{\text{ny}},3f_{\text{ny}}]\)), where frequencies have been restored to their respective Nyquist intervals.}
    \label{fig:snfscheck}
\end{figure}

As first proposed by \citet{2025A&A...693A..63W}, the sLSP method is highly effective for the identification of SNFs. Its efficacy has been demonstrated in both \(\gamma \) Dor and $\delta$~Scuti stars \citep{2025Univ...11..246Y}. To perform a comprehensive survey of SNF identification in $\delta$~Scuti stars, we developed an automated pipeline, \texttt{sLSP4SNFs}, designed to classify SNFs using a multi-parametric diagnostic framework (Section~\ref{sec:method}). Using this pipeline, we successfully identified 14,824 SNFs in 1,291 stars, extracted from a comprehensive dataset of 259,883 frequencies with SNR \(>8\) across 1,838 \textit{Kepler} $\delta$~Scuti pulsators.

\subsection{The robustness of the SNF catalog}
The confidence of our identification was evaluated through manual vetting. This involved five iterations of random sampling, where each iteration consisted of 200 SNFs: 100 with \(\mathrm{SNR_{mod}}\in [5, 10)\) and 100 with \(\mathrm{SNR_{mod}}\ge 10\). We found no misclassifications in the high-significance group (\(\mathrm{SNR_{mod}}\ge 10\)). For the intermediate group, the five trials yielded 1, 3, 2, 2, and 3 misclassified cases, respectively, resulting in a false-alarm probability (FAP) of 2.2\%. This indicates that the majority of candidates in this range are bona fide SNFs. Nevertheless, these false positives might be considered intrinsic SNFs if the selection criteria were relaxed. Figure\,\ref{fig:sLSPmissingSNF} shows an example of such a misidentified SNF, where the frequency modulation pattern closely mimics that of an intrinsic signal. The intrinsic amplitude and frequency modulation of such signals may compromise the classification accuracy of the \texttt{sLSP4SNFs} code. We note that these potential false SNFs are retained in our final catalog, as their manual removal through visual inspection would be prohibitively labor-intensive. For less significant candidates with \(\mathrm{SNR_{mod}}\in (3, 5)\), \texttt{sLSP4SNFs} achieved a validation recovery rate of 94.7\% after manual inspection (see Section~\ref{sec:method}).

The robustness of our SNF catalog was further vetted through two independent approaches. We first sought \textit{Kepler} short-cadence (SC) data for our 1,291 SNF hosts, which we found for 324 (25\%) of them. \textit{Kepler} SC data have a cadence of 58.85\,s, corresponding to a Nyquist frequency of \(\sim 8496\,\mu\mathrm{Hz}\), well above the LC Nyquist limit \citep{2010Sci...327..977B}. Of the 3,889 SNFs found in the LC data, 323 (8.3\%) were undetectable in the SC data because their amplitudes are at or below the noise level.\footnote{Typically, only one month of SC data are available, compared to four years of LC data, hence the noise is higher and the frequency resolution is poorer} in SC data. For the remaining 3,566, we confirmed 3,509 (98.4\%) common SNFs at their expected positions in the \([f_{\text{ny}},2f_{\text{ny}}]\) range. The remaining 57 (1.6\%) SNF candidates are false positives and no further attempt was made to recover them. The consistency between both comparison results confirms the reliability of the majority of our SNF candidates.

In our second validation approach, we applied the technique described by \citet{2013MNRAS.430.2986M,2019MNRAS.485.2380M}, verifying our recovered frequencies on the premise that the real frequency must possess the dominant amplitude compared to its aliases. Of the 14,824 recovered SNFs, 14,509 (97.9\%) follow this rule, aligning closely with the results of our manual vetting. Figure\,\ref{fig:snfscheck} displays the results of the amplitude comparison between the SNF aliases in the \([0,f_{\text{ny}}]\) and \([f_{\text{ny}},2f_{\text{ny}}]\) ranges, complemented by two representative examples of successful and failed identifications. Additionally, 32 SNFs were assigned to the \([3f_{\text{ny}},4f_{\text{ny}}]\) range according to the dominate amplitude. Those extra findings motivate us to combine this method with our \texttt{sLSP4SNFs} code to search for other higher SNFs beyond 2$f_\text{ny}$. Recognizing that neither approach alone was sufficient for this task, we integrated the two methods. This synergy, combined with an automated search and manual verification, ultimately revealed 207 SNFs within the \([2f_{\text{ny}}, 5f_{\text{ny}}]\) interval. Furthermore, the new method recovered 291 additional SNFs in the \([f_{\text{ny}}, 2f_{\text{ny}}]\) range that had previously missed by sLSP4SNFs, primarily due to significant residuals and the overly stringent criteria established in Section~\ref{sec:method}.

We have compiled all SNF entries in Table\,\ref{tab:snf_catalog_excerpt}, with their true frequencies recovered through the integration of these methods. Overall, Table\,\ref{tab:snf_catalog_excerpt} presents a total of 15,265 SNFs identified across a sample of 1,309 stars. 
To ensure the robustness of our catalog, we have flagged these two types of cross-identifications with specific labels in Table\,\ref{tab:snf_catalog_excerpt}. Users can regard the frequencies for which all methods agree as having the most confident labels. Nevertheless,  recognizing that the full catalog of 259,883 frequencies may still be valuable for specific research needs, it is provided in Table\,\ref{tab:all_catalog}. 

\begin{table}
\caption{All 259,883 frequencies in 1,838 $\delta$~Scuti stars. The full catalog is available in machine-readable form in the online version.}
\label{tab:all_catalog}
\centering
\begin{tabular}{c c c c c} 
\hline\hline
KIC & Frequency & $\sigma_\mathrm{f}$ & Amplitude & $\sigma_{\rm A}$ \\
 & ($\mu$Hz) & ($\mu$Hz) & (ppt) & (ppt)  \\
\hline
1026255 & 24.782977 & 0.000147 & 0.0307 & 0.0010 \\
1026255 & 49.531118 & 0.000156 & 0.0270 & 0.0010 \\
1026255 & 74.311904 & 0.000428 & 0.0091 & 0.0009 \\
1026255 & 192.306487 & 0.000107 & 0.0278 & 0.0007 \\
1573149 & 85.619497 & 0.000065 & 0.0640 & 0.0010 \\
1573149 & 171.239513 & 0.000310 & 0.0122 & 0.0009 \\
1575977 & 1.811605 & 0.000451 & 0.1635  & 0.0170 \\
\dots & \dots & \dots & \dots & \dots \\
\hline
\end{tabular}
\tablefoot{
 Frequencies are identified with SNR > 8 and serve only as reference. Any detailed asteroseismic study would require further analysis of amplitude spectra based on this table, in particular for low-amplitude signals.
}
\end{table}

\subsection{The distribution of SNFs}

The SNF distribution in $\delta$~Scuti stars follows a trend remarkably similar to that of \(\gamma \) Doradus stars, characterized by a distinct and monotonic rise in both population size and detection rate towards higher frequencies. However, since \(\gamma \) Dor stars are dominated by lower-frequency g modes and only $\gamma$~Dor--$\delta$~Sct hybrids will have a few high p modes, we observe a significantly higher number and detection rate of SNFs in $\delta$~Scuti stars, whose p modes more commonly have frequencies near the \textit{Kepler} LC Nyquist frequency. Specifically, we find a maximum rate of approximately 23\% in this study, compared to about 7\% reported for predominantly \(\gamma \) Dor stars. These results imply that any precise seismic analysis of \textit{Kepler} $\delta$~Scuti stars must carefully account for the high incidence of SNFs. Currently, as illustrated in Fig.\,\ref{fig:snf_distribution}, the detection rate shows a discontinuity around the Nyquist frequency, which we would not expect to be present if we had a 100\% recovery rate. In fact, one can see that the size of this discontinuity depends on the SNR of the observed peak, meaning that our completeness is higher for more significant modes. The sensitivity of \texttt{sLSP4SNFs} is primarily constrained by the amplitude threshold; consequently, a portion of the SNF population remains undetected because low-amplitude frequencies do not exhibit sufficient diagnostic signatures for reliable classification, lying beyond the recovery capabilities of \texttt{sLSP4SNFs} and also below the required SNR in SC data. The SC spectrum has both a higher noise level and a poorer frequency resolution, making peaks less detectable; the latter specifically complicates peak profiles in dense regions of the amplitude spectrum. In addition, the performance of \texttt{sLSP4SNFs} is very limited in dense frequency regions, thereby impeding the successful identification of SNFs. In such cases, we encourage users to apply the amplitude method as well. At this stage, identifying missing SNFs among low-amplitude signals or in dense frequency region remains a challenge, implying that we effectively provide only a lower limit for the number of intrinsic SNFs but with high confidence. 

Based on our catalog, we performed a statistical analysis examining the dependence of SNFs on stellar parameters and evolutionary stages across the HR diagram (Fig.\,\ref{fig:HR_fig}). We report no clear feature for the distribution of frequency number per star. This stands in contrast to theoretical predictions that stars in the middle of the instability strip should exhibit a higher number of excited radial modes \citep{2019MNRAS.490.4040A}. The caveat is that only frequencies with relatively high amplitudes were extracted in this work. 
However, we do observe a clear trend that younger $\delta$~Scuti stars exhibit a higher fraction of SNFs relative to their total frequency count. This distribution is similar to the regular frequency patterns in young $\delta$~Scuti stars as identified by large frequency separation $\Delta\nu$ \citep{2020Natur.581..147B}. The SNF fraction gradually decreases as the stars evolve, a result consistent with the median values of our theoretical eigenfrequencies calculated by GYRE. This can be understood in the context of stellar evolution: as stars evolve off the ZAMS, the mean density $\bar\rho$ decreases, causing the eigenfrequencies of a specific p mode (with fixed \(n\), \(\ell \) and \(m\)) to shift toward lower values \citep{2026MNRAS.545f2001G}, as defined by the pulsation constant:
\begin{equation}
Q = P \sqrt{\frac{\bar{\rho}}{\bar{\rho}_{\odot}}} \sim \frac{\Delta\nu}{\nu},
\end{equation}
where \(P\) is the pulsation period, \(\bar{\rho }_{\odot }\) represents the mean solar density, $\nu$ is the frequency value and $\Delta\nu\propto ({\bar{\rho}})^{0.5}$. Consequently, the representative frequencies are less likely to exceed \(f_{\text{ny}}\). 
Therefore, both the SNF count and its corresponding fraction, as a proxy of higher frequencies, decrease as $\delta$~Scuti stars evolve away from the ZAMS phase, which is consistent with the results from both observation and stellar models \citep{2020Natur.581..147B,2023MNRAS.526.3779M,2026MNRAS.545f2001G}.

\subsection{Prospects}
Our observational results demonstrate that any mode identification in $\delta$~Scuti stars should be approached with caution when the sampling cadence approaches or exceeds 30 minutes. The recovery of SNFs can alter the existing relationships among pulsation modes, such as linear combination frequencies, which exhibit a \(\sim\)17\% discrepancy in our catalog. This subsequently biases the characterization of internal stellar properties derived from seismic models, especially if an independent mode is misidentified as a linear combination, or vice versa. In this context, seismic analysis of $\delta$~Scuti stars closer to the terminal-age main sequence (TAMS) appears more reliable than that of younger MS stars, as the probability of SNF pollution is significantly reduced. 

Our catalog offers users an easy method to check whether an observed peak is a Nyquist alias, and provides its real frequency if so. It also provides information on the SNR of the modulation, as a proxy of confidence in the identification. For specific stars, to improve the accuracy of SNF identification, we combined our method with amplitude ratio between aliases and cross-checking available \textit{Kepler} SC data. We also encourage the use of other independent techniques for further validation, such as analyzing amplitude modulation \citep{2021RNAAS...5...41Z} and identifying equidistant frequency multiplets \citep{2013MNRAS.430.2986M}. The most appropriate method will depend on what data are available, but ideally one can apply multiple methods and arrive at the same conclusions. This workflow will increase the reliability of asteroseismic analyses of delta Scuti stars in the \textit{Kepler} LC data. 

Our method is based on the principle that a slight correction to the sampling time introduces periodic frequency modulations with a predictable pattern. \citet{2014MNRAS.441.2515M} showed that, in some cases, their signatures may be misinterpreted as systematic phase modulations induced by binarity, especially considering the similarity between SNF modulation and binary modulation with periods around one year. To distinguish between them, one must scrutinize all significant frequencies. Phase modulation from binarity exhibits the same pattern across all significant frequencies and can modulate on various scales as determined by orbital parameters, whereas SNF modulations are restricted to aliased frequencies originating from the super-Nyquist regime with a predicted frequency scale and phase with respect to \textit{Kepler}'s heliocentric orbit. We finally stress that SNFs exhibit a characteristic modulation pattern that is clearly distinguishable from the diverse intrinsic variations of nonlinear mode interactions in different types of pulsators \citep[see, e.g.,][]{2016MNRAS.460.1970B,2016A&A...585A..22Z}. However, as shown in Fig.~\ref{fig:sLSPmissingSNF}, the observed modulation can sometimes be a composite of the SNF pattern and intrinsic nonlinear resonance, which complicates the identification of individual SNF frequencies.

We end this paper with a brief summary: we have established a comprehensive catalog of high-confidence SNFs for \textit{Kepler} $\delta$~Scuti stars, demonstrating that SNFs are a ubiquitous feature in high-frequency pulsators observed with \textit{Kepler} long-cadence photometry. This catalog provides a foundation for future asteroseismic investigations of $\delta$~Scuti stars. It enables future studies to re-examine previous mode identifications and, based on our findings, to identify regular frequency spacings in young stars and re-evaluate linear combination frequencies. Our findings clearly reveal that SNF modulation patterns are distinct from those induced by intrinsic nonlinear resonances or binary orbital effects, providing a robust method to disentangle these phenomena.

\begin{acknowledgements}
      The authors acknowledge the support from the National Natural Science Foundation of China (NSFC) through grant Nos. 12273002, 12541303 and 12427804, and the Central Guidance for Local Science and Technology Development Fund under No. ZYYD2025QY27. This work is partially supported by the science research grants from the China Manned Space Project. SJM was supported by the Australian Research Council through Future Fellowship FT210100485. S.C. acknowledges support from the Centre National d’Etudes Spatiales (CNES, France),  focused on the {Kepler} Mission. All of the \textit{Kepler} data used in this paper can be found in MAST. The authors appreciate all who have contributed to making these missions possible. Funding for the {Kepler} Mission is provided by NASA’s Science Mission Directorate.
\end{acknowledgements}

\bibliographystyle{aa} 
\bibliography{references}

\end{document}